\documentclass[9pts,conference,hidelinks]{IEEEtran}
\IEEEoverridecommandlockouts
\usepackage{cite}
\usepackage{amsmath,amssymb,amsfonts}
\usepackage{graphicx}
\usepackage{textcomp}
\usepackage{balance}
\usepackage{xcolor}
\usepackage{algorithm}
\usepackage[noend]{algpseudocode}
\usepackage{mathtools}
\usepackage{multirow}
\usepackage{enumitem}
\usepackage{caption}
\usepackage{subcaption}
\usepackage{calc}
\usepackage{wrapfig}
\usepackage{comment}
\usepackage{etoolbox}
\usepackage{booktabs}
\usepackage{adjustbox}
\usepackage{listings}
\usepackage{xcolor}
\usepackage[colorinlistoftodos]{todonotes}
\usepackage{pgfplots}
\usepackage{pgfplotstable}

\def\baselinestretch{0.93}

\usepackage[colorlinks]{hyperref}
\hypersetup{
    colorlinks = true,
    filecolor = magenta,
    urlcolor = magenta,
    citecolor = cyan,
    linkcolor = magenta,
}

\usepackage{url}

\usepackage{breakurl}

\definecolor{codegreen}{rgb}{0,0.6,0}
\definecolor{codegray}{rgb}{0.5,0.5,0.5}
\definecolor{codepurple}{rgb}{0.58,0,0.82}
\definecolor{backcolour}{rgb}{0.95,0.95,0.92}

\lstdefinestyle{mystyle}{
    backgroundcolor=\color{backcolour},   
    commentstyle=\color{codegreen},
    keywordstyle=\color{magenta},
    numberstyle=\tiny\color{codegray},
    stringstyle=\color{codepurple},
    basicstyle=\ttfamily\footnotesize,
    breakatwhitespace=false,         
    breaklines=true,                 
    captionpos=b,                    
    keepspaces=true,                 
    numbers=left,                    
    numbersep=5pt,                  
    showspaces=false,                
    showstringspaces=false,
    showtabs=false,                  
    tabsize=2,
    moredelim=**[is][\color{red}]{@}{@}1
}
\def\BibTeX{{\rm B\kern-.05em{\sc i\kern-.025em b}\kern-.08em
    T\kern-.1667em\lower.7ex\hbox{E}\kern-.125emX}}

\usepackage{siunitx}

\usepackage{xcolor}
\usepackage{ifthen}
\usepackage{mathabx}
\newboolean{showcomments}
\setboolean{showcomments}{true}
\ifthenelse{\boolean{showcomments}}
{ \newcommand{\mynote}[3]{
     \fbox{\bfseries\sffamily\scriptsize#1}
        {\small$\blacktriangleright$\textsf{\emph{\color{#3}{#2}}}$\blacktriangleleft$}}}
{ \newcommand{\mynote}[3]{}}
\definecolor{asparagus}{rgb}{0.53, 0.66, 0.42}

\begin{document}

\title{\LARGE 
REFRESH FPGAs: Sustainable FPGA Chiplet Architectures 
\vspace{-8pt}
}

\makeatletter
\patchcmd{\@maketitle}
  {\addvspace{0.5\baselineskip}\egroup}
  {\addvspace{-1.0\baselineskip}\egroup}
  {}
  {}
\makeatother

\author{\IEEEauthorblockN{Peipei Zhou\IEEEauthorrefmark{1}, Jinming Zhuang\IEEEauthorrefmark{1}, Stephen Cahoon\IEEEauthorrefmark{1}, Yue Tang\IEEEauthorrefmark{1},\\ Zhuoping Yang\IEEEauthorrefmark{1}, Xingzhen Chen\IEEEauthorrefmark{1}, 
Yiyu Shi\IEEEauthorrefmark{2}, Jingtong Hu\IEEEauthorrefmark{1}, 
Alex K. Jones\IEEEauthorrefmark{1}}
\IEEEauthorblockA{\IEEEauthorrefmark{2}University of Notre Dame,
yshi4@nd.edu}
\IEEEauthorblockA{\IEEEauthorrefmark{1}University of Pittsburgh,
\{peipei.zhou, jthu, akjones\}@pitt.edu}
}

\maketitle



\begin{abstract}

\noindent \textbf{}
There is a growing call for greater amounts of increasingly agile computational power for edge and cloud infrastructure to serve the computationally complex needs of ubiquitous computing devices. Thus, an important challenge is addressing the holistic environmental impacts of these next-generation computing systems. To accomplish this, a life-cycle view of sustainability for computing advancements is necessary 
to reduce environmental impacts such as 
greenhouse warming gas emissions from these computing choices.  
Unfortunately, decadal efforts to address operational energy efficiency in computing devices have ignored and in some cases exacerbated embodied impacts from manufacturing these edge and cloud systems, particularly their integrated circuits.    During this time FPGA architectures have not changed dramatically except to increase in size. 
Given this context, we propose REFRESH FPGAs to build new FPGA devices and architectures from recently retired FPGA dies using 2.5D integration.  To build REFRESH FPGAs requires creative architectures that leverage existing chiplet pins with an inexpensive to-manufacture interposer coupled with creative design automation.  In this paper, we discuss how REFRESH FPGAs can leverage industry trends for renewable energy integration into data centers while providing an overall improvement for sustainability and amortizing their significant embodied cost investment over a much
longer ``first'' lifetime.  

\end{abstract}

\section{Introduction}\label{sec:introduction}
\vspace{-.05in}

As we have become firmly ensconced in the post Moore era, computer architectures have turned to accelerators for executing computationally and/or memory intensive applications with improved performance. However, a new emerging concern is the environmental impacts of decisions made about these next generation architectures.  Until recently, sustainable computing was highly concerned with operational energy efficiency to reduce greenhouse warming gas (GHG) emissions, such as CO$_2$, from electricity generated with fossil fuels to power these systems.  However, continuing advances in renewable energy integration, coupled with the realization of the significant and in many cases, dominant, embodied GHG emissions from chip manufacturing for these systems, has changed the calculus of sustainable computing~\cite{Gupta-HPCA21-Chasing,KLINE2019322,ICCAD13-Jones,Dark-Silicon-Harmful,lieven-cal-model}.  

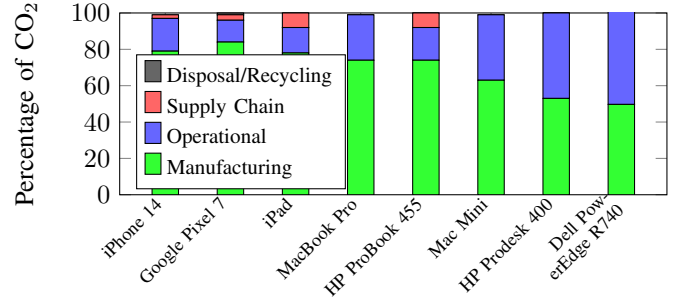
\begin{figure}
\centering
    \pgfplotstableread{
Label                   Manufacturing   Operational SupplyChain Disposal 
{iPhone 14}             79              18          2           0
{Google Pixel 7}        84              12          3           1    
iPad                    78              14          8           0
{MacBook Pro}           74              25          0           0
{HP ProBook 455}        74              18          8           0
{Mac Mini}              63              36          0           0
{HP Prodesk 400}        53              47          0           0
{Dell PowerEdge R740}   49.7            52.5        0           0

    }\testdata

    \begin{tikzpicture}
    \begin{axis}[
        height=4cm,
        width=\columnwidth,
        ybar stacked,
        ymin=0,
        ymax=100,
        xtick=data,
        ylabel={Percentage of CO$_2$},
        legend style={cells={anchor=west}, legend pos=south west, font=\footnotesize},
        reverse legend=true, 
        xticklabels from table={\testdata}{Label},
        xticklabel style={font=\scriptsize,rotate=45,text width=2cm,align=right,anchor=east}, 
    ]
    \addplot [fill=green!80] table [y=Manufacturing, meta=Label, x expr=\coordindex] {\testdata};
    \addlegendentry{Manufacturing}
    \addplot [fill=blue!60] table [y=Operational, meta=Label, x expr=\coordindex] {\testdata};
    \addlegendentry{Operational}
    \addplot [fill=red!60] table [y=SupplyChain, meta=Label, x expr=\coordindex] {\testdata};
    \addlegendentry{Supply Chain}
    \addplot [fill=black!60] table [y=Disposal, meta=Label, x expr=\coordindex] {\testdata};
    \addlegendentry{Disposal/Recycling}


    \end{axis}
    \end{tikzpicture}
\vspace{-40pt}
\caption{Sources of CO$_2$e from different computing products.}
\label{fig:CO2-sources}
\vspace{-15pt}
\end{figure}

Fig.~\ref{fig:CO2-sources} shows examples of lifecycle assessments of a variety of computing devices.  Mobile devices tend to exceed 75\% of their carbon as embodied carbon.  However, even desktop and server machine examples show at least 50\% of their carbon from embodied carbon.  
For data center systems, the \textit{embodied} energy was about 33\% of the overall energy compared to 65\% from operational energy.  However, with renewable integration, the embodied \textit{carbon} is 82\% compared to 18\% operational \textit{carbon} emissions in leading hyper scalars~\cite{Gupta-HPCA21-Chasing,DJslides}.  
Thus, there is a growing movement to address environmental impacts, holistically, for computing systems throughout all stages of their life-cycle, including manufacturing, supply chain, operation, and disposal. \textit{Focusing solely on operational energy-efficiency without taking the embodied environmental impacts of computing systems into consideration cannot achieve true sustainability in the long run.}

\begin{figure*}
\begin{center}
\includegraphics[width=\textwidth]{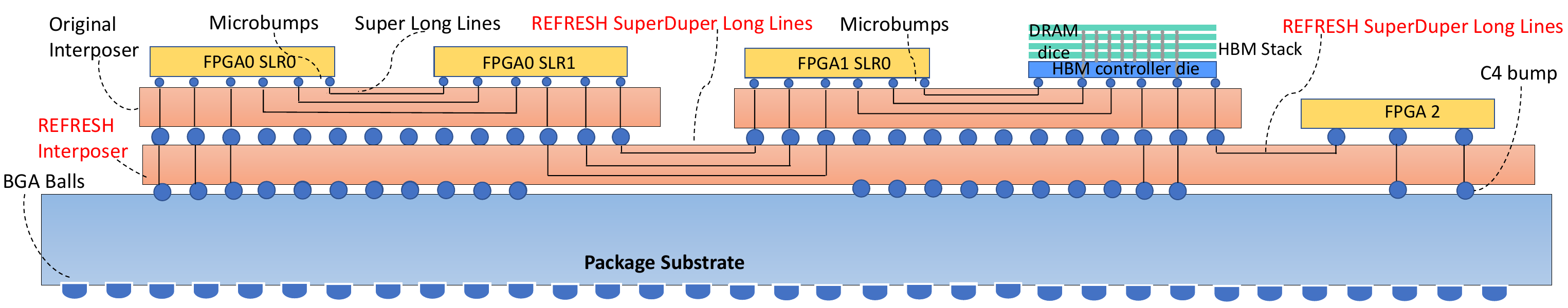}
\caption {REFRESH interposer for integration of homogeneous and heterogeneous monolithic and/or chiplet-based FPGAs.  
}
\label{fig:refresh_fpga}
\end{center}
\vspace{-0.3in}
\end{figure*}


Towards this goal, we propose REFRESH or \textit{Revisiting Expanding FPGA Real-estate for Environmentally Sustainable Heterogeneous-Systems}. 
The REFRESH concept is based on several FPGA-specific observations. The replacement cycle for systems with accelerators (particularly FPGAs) is very fast (circa two years)~\cite{TechTarget} due to increasingly short support lifetimes by the vendors. 
However, these retired devices have many years of effective service life remaining.  For sustainability, amortizing their embodied environmental impact investment over a longer service life is desirable.  Moreover, the regularity, maturity, and flexibility of these devices suggest they have the most potential for obtaining value from increasingly long lifetimes.  Furthermore, REFRESH reduces pressure to extract raw materials such as rare earth minerals, while also significantly reducing the growing environmental risks of e-waste by keeping these toxic, non-biodegradable devices out of landfills and reducing their negative impacts accumulating in the soil, air, water minimizing health impacts to living things.

REFRESH proposes to build \textbf{``new'' FPGA devices} from recently retired FPGAs using 2.5D integration of these FPGA dies with an underlying interposer.  This allows for an interconnection between FPGA chiplets as well as thru silicon vias (TSV) to the underlying package pins with an example in Fig.~\ref{fig:refresh_fpga}.  
This allows FPGA devices to achieve a much longer ``first'' lifetime while meeting the needs of accelerator programmers to provide increasingly large and capable configurable fabrics.  
Leveraging the increasing investment of renewable energy, \textit{moderate increases in operational energy may have a minimal negative environmental impact while achieving substantial improvements in embodied environmental impacts.}  

\vspace{-.05in}
\section{REFRESH Concept and Sustainability}\label{sec:architecture}
\vspace{-.05in}

Modern FPGAs are already transitioning to chiplet-based design to increase yields for such large devices.  Building REFRESH FPGAs requires advances to FPGA architecture and design flows that are consistent with the challenges of designing for chiplet-based FPGAs with new challenges of more restricted long distance interconnect as well as challenges of reliability for which FPGA architectures are well suited.

\vspace{-2pt}
\subsection{REFRESH Architecture and Design Automation Co-Design}
\vspace{-3pt}
Chiplet-based FPGAs must address the limited bandwidth between chiplets, often referred to as \textit{super long lines} (SLLs).  Because dies for REFRESH devices have already been packaged, in REFRESH we introduce the concept of \textit{super \textbf{duper} long lines} (SDLLs) to account for I/O that has already been routed to package pins.  In Fig.~\ref{fig:refresh_fpga} we show that REFRESH devices may integrate monolithic devices as well as chiplet-based devices.  Thus, characterization of communication (SLLs and SDLLs) across boundaries through an interposer is critical to inform chiplet layout and interposer design for REFRESH FPGAs including consideration of which homogeneous and heterogeneous architectures, potentially including devices from different generations and high-bandwidth memory can be retrofitted into 2.5D System-in-Package (SiP) design.  
A critical tool for programming these devices will be a fine grain automated flow to partition designs across chip boundaries~\cite{cong2018latte,guo2021autobridge,guo2020analysis,rapidstream2022fpga,rapidstream2023TRETS}. 

\setlength{\tabcolsep}{6pt}
\begin{table}[bp]
\vspace{-5pt}
\caption{32-bit floating-point matrix multiplication implemented on different FPGA generations.}
\vspace{-.1in}
\begin{tabular}{l|rcrrr}
\hline\hline
\textbf{Class}                         & \textbf{Tech.} & \textbf{Board /} & \multicolumn{1}{c}{\textbf{Latency}} & \multicolumn{1}{c}{\textbf{Dynamic}} & \multicolumn{1}{c}{\textbf{Static}} \\
& \textbf{Node} &  \textbf{Device}
& 
& \multicolumn{1}{c}{\textbf{Power}}
& \multicolumn{1}{c}{\textbf{Power}}
\\\hline
\textbf{Virtex-7} & \SI{28}{\nano \meter}               & VC709                 & {\color[HTML]{333333} \SI{6.09}{\nano \second}}          & {\color[HTML]{333333} \SI{21.835}{\watt}}              & {\color[HTML]{333333} \SI{0.799}{\watt}}              \\
\textbf{Ultrascale+}    & \SI{16}{\nano \meter}               & ZCU102                & {\color[HTML]{333333} \SI{4.60}{\nano \second}}          & {\color[HTML]{333333} \SI{21.410}{\watt}}               & {\color[HTML]{333333} \SI{0.920}{\watt}}               \\
\textbf{Versal}               & \SI{7}{\nano \meter}                & VMK180                & {\color[HTML]{333333} \SI{3.99}{\nano \second}}             & {\color[HTML]{333333} \SI{12.738}{\watt}}              & {\color[HTML]{333333} \SI{9.384}{\watt}} \\            
\hline\hline
\end{tabular}
\label{tab:matmul}
\end{table}

\vspace{-5pt}
\subsection{REFRESH Hardware Analysis and Conceptualization}
\vspace{-3pt}
FPGA architecture has been relatively static in terms of innovation in look-up tables, multiply accumulate units, block memories, etc., over the last decade or more.  Actual FPGA advances are in the capacity of what FPGAs can support while the performance in terms of clock frequency and energy for a fixed design has not improved dramatically~\cite{FPGA-Survey-21}.

To demonstrate this, we implemented a 32-bit floating point matrix multiplication design on three generations of FPGA fabrics from AMD/Xilinx, shown in Table~\ref{tab:matmul}.  Each new generation benefits, however the improvement is not dramatic (50\% improvement in latency from \SI{28}{\nano \meter} to \SI{7}{\nano \meter}).  The dynamic power drops from \SI{22}{\watt} to \SI{13}{\watt}, but the static power grows from \SI{0.8}{\watt} to almost \SI{10}{\watt}.  These power estimates are from the AMD/Xilinx tool flow.  

\setlength{\textfloatsep}{10pt}
\begin{figure}[tbp]
     \subfloat[Fabrication in\\ Taiwan used in California]{
     \includegraphics[height=1.25in]{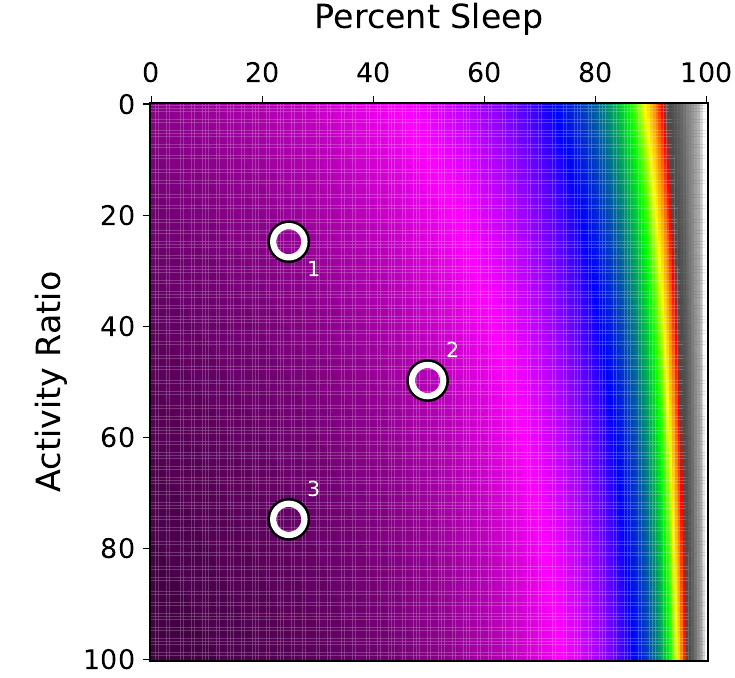}
     \label{fig:tai-ca}
     } \hfill
     \subfloat[Fabrication in Taiwan used with 90\% renewables] {
     \includegraphics[height=1.25in]{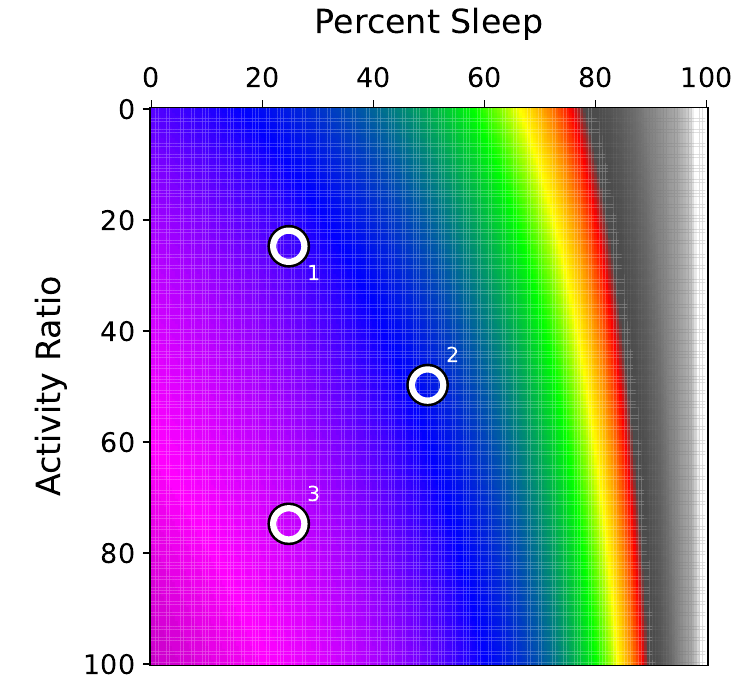}    
     \label{fig:tai-rew}
     } \hfill
     \includegraphics[height=1.25in]{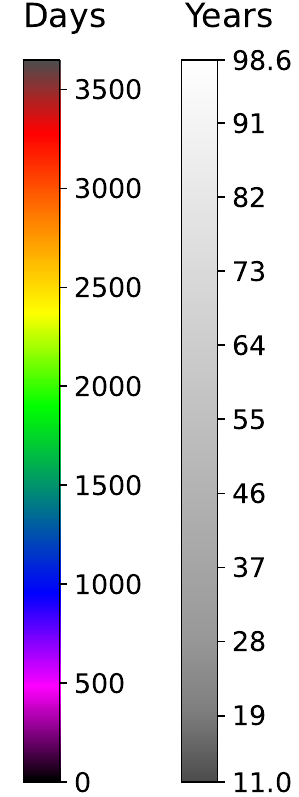}
    
     \caption{Carbon indifference plots for a VM1802 vs. a REFRESH FPGA made from four ZCU102 2.5D integrated dies.}
     \label{fig:impact-of-REFRESH}
     \vspace{-.1in}
 \end{figure}
To compare two design choices for their sustainability requires we combine the contribution of manufacturing (embodied) impacts ($E_i$) and operational impacts ($O_i$) of such systems into a relevant number based on the system lifetime ($L_i$). 
We do this by using indifference and break-even 
calculations~\cite{KLINE2019322} as described in Eq.~\ref{eq:indiff}.
\vspace{-3pt}
\begin{equation}
t_I \!=\! \frac{E_1 - E_0}{(O_0\!+\!\frac{E_0}{L_0})\!-\!(O_1\!+\!\frac{E_1}{L_1})} \; t_B \!=\! \frac{E_1}{(O_0\!+\!\frac{E_0}{L_0})\!-\!(O_1\!+\!\frac{E_1}{L_1})}
\label{eq:indiff}
\end{equation}
To demonstrate the potential value of REFRESH FPGAs we show some system-level comparisons in Figure~\ref{fig:impact-of-REFRESH}.  First in Figure~\ref{fig:tai-ca}, we show the indifference point comparison ($t_I$) of large-scale matrix multiplication using a VM1802 FPGA compared to a first-order approximation of a REFRESH FPGA comprised of four ZCU102 devices. 
The VM1802 has a significantly higher embodied contribution than the REFRESH device, with the REFRESH device having a higher operational contribution with a lower performance.  Thus $t_I$ is when the VM1802 saves enough operational carbon to meet the REFRESH FPGA.  

We show three cases: $r_\text{sleep}=\{25\%,50\%,25\%\}$, $r_\text{active}=\{25\%,50\%,75\%\}$ such that $r_\text{sleep}$ is the sleep time to total time in service and $r_\text{active}$ is the computation time  versus non-sleep time, including idle time.  
Cases 1 and 2 do a similar amount of work but have different sleep-to-idle ratios, while case 3 does 3$\times$ the work of Case 1.
The VM1802 FPGA accelerator fabricated in Taiwan and used in CA shows that it has a $t_I$ of $\leq1$ year because the operational savings eventually makes up for the embodied overhead.  However, as renewable energy penetration increases, the indifference time increases, reaching three years for cases 1 and 2, and two years for case 3.

\section{Conclusions}
Several critical challenges remain to be solved to build effective and reliable REFRESH devices including addressing the impact of die aging, die connection topology, connection bandwidth, architectural choices, fault tolerance, and replacement cycles,  for current and future acceleration workloads.
However, REFRESH FPGAs have the potential to provide better sustainability over the system lifecycle~\cite{ICCAD13-Jones,6604497,Dark-Silicon-Harmful,10.1145/3060403.3066859,KLINE2019322,7892605,9941196}
for applications such as hyperdimensional computing, deep learning~\cite{10.1145/3489517.3530509,fpga23acap,dac23acap,CORUSCANT,POD-RACING,zhang2021algorithm,cong2018latte,zhou2016energy,zhang2018caffeine}, and bioinformatics~\cite{zhou2021mocha,lo2020algorithm,zhou2018doppio}.

\balance

\bibliographystyle{IEEEtran}
\bibliography{bib/ref}


\end{document}